\documentclass[10pt,letterpaper]{article} 
\usepackage[english]{babel}
\newcommand{\lb}{\langle}
\newcommand{\rb}{\rangle}

\newcommand{\beq}{\begin{equation}}
\newcommand{\eeq}{\end{equation}}
\newcommand{\lbl}{\label}
\newcommand{\beqnar}{\begin{eqnarray}}
\newcommand{\eeqnar}{\end{eqnarray}}
\newcommand{\beqnars}{\begin{eqnarray*}}
\newcommand{\eeqnars}{\end{eqnarray*}}

\newcommand{\goesto}{\rightarrow}
\newcommand{\s}{\\[1ex]}
\newcommand{\re}[1]{(\ref{#1})}
\newcommand{\q}{\quad}

\newcommand{\tnr}{\otimes} 
\newcommand{\tr}{\mbox{Tr }}
\newcommand{\trM}{\mbox{Tr}_M\,} 
\newcommand{\trS}{\mbox{Tr}_S\,} 
\newcommand{\dajA}{\hat{\bf A}}
\newcommand{\dajE}{\hat{\bf E}}
\newcommand{\dajrho}{\hat{\bf \rho}}
\newcommand{\dajEj}{\hat{{\bf E}_j}} 
\newcommand{\dajM}{\hat{{\bf M}}}
\newcommand{\dajMj}{\hat{{\bf M}}_j}
\newcommand{\dajUj}{\hat{\bf U}_j}
\newcommand{\dajU}{{\bf U}}
\newcommand{\dajV}{{\bf V}}
\newcommand{\dajF}{{\bf F}}
 
\newcommand{\dajGj}{\hat{{\bf G}_j}}
\newcommand{\dajPij}{\hat{{\bf \Pi}_j}}
\newcommand{\dajPik}{\hat{{\bf \Pi}_k}}
\newcounter{myeq}
\newcommand{\calM}{{\cal M}}
\newcommand{\nullsp}{\mbox{Null}}
\newcommand{\init}{\mbox{In}}
\newcommand{\range}{\mbox{Range}}

\newcounter{saveeq}

\begin{document} 
%
%\begin{center}
%\Large
%``Contextual weak values'' of quantum measurements are not limited
%to the traditional weak value
%\\[1ex] 
%\ 
%\normalsize
%by 
%\\[1ex] 
%\large 
%Stephen Parrott 
%\normalsize
%\\[1ex]
%February 20, 2011 
%\end{center}
%\noindent 
\title{
%\large 
``Contextual weak values'' of quantum measurements with positive
measurement operators are not limited
to the traditional weak value
\\[1ex]
\normalsize
%by
%\\[1ex]
%\large
%J.\ Dressel, S.\ Agarwal, and A.\ N.\ Jordan.
} 
%\footnote{Running head:} 
\author{Stephen Parrott\thanks{For contact information, 
go to http://www.math.umb.edu/$\sim$sp}}
%} 
\date{May 2, 2011}
\maketitle
\begin{abstract}
A recent Letter in Physical Review Letters, ``Contextual Values of Observables
in Quantum Measurements'', by J. Dressel, S. Agarwal, and A. N. Jordan
\cite{DAJ}
(abbreviated DAJ below), 
introduces the concept of ``contextual values'' and claims that it leads to
``a natural definition of a general conditioned average that converges
uniquely to the quantum weak value in the minimal disturbance limit''.
However, they do not define ``minimal disturbance limit''.
The present paper is in part the saga of my search for a definition
of ``minimal disturbance limit'' under which this claim could be proved.
The search finally ended with what is probably a definitive counterexample to the
claim. 

In addition, the present work analyzes the arguments
of DAJ in detail and relates them to traditional weak value theory.  Various
gaps and possible errors in its arguments are explicitly noted.
\end{abstract} 
%
%\noindent
%\ 
%\\
%{\bf \large DRAFT: Please do not circulate}
%
\section{Introduction for Versions 4 and 5}
This Introduction assumes that the reader is at least vaguely familiar
with Versions 2 and 3.  The next section gives a more leisurely introduction 
reprinted from Version 2.

This paper has gone through a number of incarnations.  Version 1 and
the nearly identical Version 2 formulated what seemed at the time
a reasonable definition of DAJ's%
\footnote{Abbreviation for Dressel, Agarwal, and Jordan's paper 
\cite{DAJ}.}
undefined ``minimal disturbance limit'' and presented a counterexample
to its claim that their ``general conditioned average $\ldots$
converges uniquely to the quantum weak value in the minimal disturbance
limit'', assuming this definition. 

Later, I learned that the my definition did not correspond to the authors'.
(For the strange story of how I learned this, see the "Afterword" section
below.)   My definition corresponded to what would typically be called 
``weak'' measurements, which are not the same as the ``minimally disturbing
measurements'' discussed just below.

DAJ was assuming that the quantum measurements were slightly
more general than what the   
the book of Wiseman and Milburn (\cite{wiseman}, Section 1.4.2) 
calls ``minimally disturbing measurements''.  
These are defined as measurements for which the measurement
operators are positive.  (They do not correspond to what one might
guess given only the 
normal associations of the English phrase ``minimally disturbing
measurements''; see the Afterword to Version 3 for more information.)

There is no limit in this definition, so DAJ's 
``minimal disturbance limit'' remains undefined, but I assume it is
what would normally be called a ``weak limit'' (i.e., in the limit, 
the effect of the measurement on the state being measured goes to zero)
of their generalization of Wiseman/Milburn's ``minimally disturbing''
measurements.%
\footnote{The reader is probably wondering why I don't simply ask the 
authors for their definition of ``minimal disturbance limit''.  I have 
tried repeatedly, but they ignore my inquiries.  See the Afterword 
to Version 3 (reprinted below) for more information.} 

The present Version 4 presents a counterexample to the above claim of DAJ
assuming that the measurements are ``weak'' in the sense described
in Versions 1 and 2 (i.e., what Versions 1 and 2 assumed DAJ meant
by ``minimal disturbance limit'') and in addition are ``minimally disturbing
measurements'' in the sense of Wiseman/Milburn (\cite{wiseman} just described. 

This new counterexample is in Section \ref{sec: V4example} 
and should not be confused with
the counterexample of earlier versions in Section \ref{sec: nontrad}.
Since the previous counterexample is still correct as stated, I have 
not removed it, but its hypotheses are different from those of the new
example.  Sections 2 through 11 are almost identical to Version 3; only
one significant typo has been detected and corrected. 

Version 5 is mathematically identical to Version 4.  Minor improvements
in the exposition have been made, and a bad typo in equation \re{eq460}
has been corrected. 
\section{Brief introduction from Version 2} 
This work discusses the paper 
\begin{center}
\large
 ``Contextual Values of Observables in Quantum Measurements''
\\
\normalsize
by
J.\ Dressel, S.\ Agarwal, and A.\ N.\ Jordan,
\end{center} 
which will henceforth be abbreviated DAJ.
It introduces ideas which I found very interesting.  I also found
them quite puzzling because it seemed that they might be in some way 
contradictory to the results of my recent work \cite{parrott1}
\cite{parrott2} and to prior and overlapping work of 
Jozsa \cite{jozsa}. (DAJ cites all of these works, 
but does not note the possible contradiction.)  

Every prudent author knows how easily mistakes are made and devises
tests for the internal consistency of his%
\footnote{or her, {\em of course}.  I adhere to the long-standing and
sensible grammatical convention that in contexts like this, 
``his'', ``hers'', ``his or hers'', or ``hers or his'' all carry the 
same meaning.}
work.  This note is largely a product of a search for a definite
contradiction between DAJ and my work. 
A definite contradiction eluded me because 
DAJ does not state its hypotheses in a mathematically explicit way.
However, for a period of weeks I had the uncomfortable feeling 
that a contradiction might be ``right around the corner'', 
invisible now but ready to pop into
being as soon as I sorted out the implicit assumptions of DAJ.  

I found a number of  mathematical errors in the argument of DAJ, 
but none of these seemed crucial. 
I worried that with appropriate definitions and
qualifications, the main thrust of DAJ's argument might survive to produce
a contradiction.   

Finally, I focused on one step in DAJ's argument and decided that not only
did I not know how to justify this step, but that  DAJ's formulation of
the step seemed so vague that it should be considered a major gap in their 
proof.  
It would be foolish for me to claim that there is no way to fill
this gap, but I would be very surprised if it could be filled so as 
to obtain their conclusions under reasonably general hypotheses.

This work will attempt to present the argument of DAJ in a straightforward
way which avoids some of their unnecessary and distracting subarguments,
with the aim of clearly exposing the gaps in it.  
I leave it to the reader to formulate hypotheses under 
which the gaps can be filled.  

The conclusion of their paper which initially interested me most was
their equation (7), which 
states that the ``traditional'' weak value 
\setcounter{equation}{99}
\beq
\lbl{eq1}
A_w = \Re 
\frac{\lb \psi_f |  \hat{\bf A} | \psi_i \rb}{\lb \psi_f | \psi_i \rb}
\,
\eeq 
is ``uniquely defined'' by DAJ's theory of ``contextual values'' to be
essentially this expression (more precisely, 
a generalization of this expression to mixed states, their equation (7)),
in some undefined ``minimal disturbance limit''.%
\footnote{Equation (100) is the real part of DAJ's equation (1).  
Those familiar with the ``weak value'' literature will probably
immediately guess the meaning of the undefined symbols, which are also
undefined in DAJ. $\psi_i$ is an initial state in which the expectation
of an observable ${\bf \hat{A}}$ is to be measured, and $\psi_f$ is a 
``postselected'' final state.  For more extensive definitions, see 
\cite{parrott1}.
}
This reinforces a general impression given by much of the ``weak value''
literature that the traditional weak value \re{eq1} is somehow inevitable. 

It seemed to me that the correspondence between DAJ's theory of ``contextual
values''  and traditional ``weak measurement'' theory was close enough that
if \re{eq1} were ``unique'' in the DAJ theory, then it ought to also be unique
in ``weak measurement'' theory, but I believed I had proved via 
rigorous mathematics \cite{parrott1} that the latter was false.   
I couldn't rest until I had sorted out the correspondence between 
the two theories with the object of resolving the potential contradiction.

Part of this paper explores the correspondence.  I regard this as
its more interesting and fundamental contribution.  Along the way, the
potential contradiction was resolved.
\section{Acknowledgment}

\noindent
{\bf \large The following is reprinted from Version 2:}
\\[1ex]
Acknowledgments are usually placed at the end of 
a paper, but this one is being placed at the beginning because it 
may affect how readers view various statements in it.  I sent a
draft of the paper to the authors of 
DAJ asking for their comments, with the hope of 
correcting any errors or misunderstandings.  They sent me several helpful
replies, for which I thank them.

All of the replies were prominently labeled ``Private Comments''.  Because
of this, I don't feel at liberty to quote them, nor attribute them 
to the authors in paraphrased form. 

The original draft contained a statement that their equation (4) should be
considered as a {\em hypothesis} for their (6), rather than the assertion
that the text of DAJ seems to suggest.
They confirmed that this was their intention, and specifically authorized me 
to state that, so I  have.  As of this writing, they have not specifically
authorized me to divulge anything else, so I haven't.  

However, they did give me quite a bit of information which mainly confirmed what
I had conjectured when I sent them the draft.  Consequently, this version
is little changed from the original draft.  In particular, I still believe 
that the gaps mentioned above are substantial.  

Everything that I know is incorporated in the present version, though 
I cannot attribute specific statements to the authors.  It may be 
helpful to readers to understand that I know more than I can publicly 
say. 
\\[1ex]

\noindent
{\bf\large  The following is added in Version 3:}
\\[1ex]

Since Version 2 was placed in the arXiv, significant new information
about DAJ's definition of ``minimal disturbance limit'' has surfaced.
This is presented in an ``Afterword'', which the reader may want to scan
before proceeding.  In summary, they are using their own definition of 
this term which is not fully given in DAJ and
does not correspond to the usual English meanings of ``minimal'' 
and ``disturbance''.  In particular, their concept of 
``minimal disturbance limit'' is distinct from the usual concept of 
``weak limit''. 

To reflect this, the title of Version 2 has been changed and 
its abstract rewritten.   
The body of Version 2 
is still essentially accurate, 
and has been altered only by correction of typos 
and a few potentially misleading phrases.    

The sentence
\begin{quote}
``It may be helpful to readers to understand that I know more than I can
publicly say''
\end{quote}
referred to the fact that an attempted proof that they had sent me for a 
major claim of DAJ (that their (6) implies (7) in ``the minimal
disturbance limit'') was wrong.  Up to that point, 
I had made no promises of confidentiality,
so that there is no reason that I could not have revealed that, but I felt
that I should not.
Perhaps given time they could correct the proof.   

But the situation has changed.
Months have passed, and they have ignored direct inquiries asking
if they
still believe they have a valid proof that (6) implies (7) in their 
``minimal disturbance limit''.  
I advise anyone who may be thinking of 
investing time in building on the work of DAJ to first seek a proof 
from the authors and judge for themselves.  
\section{General introduction and remarks on notation}
I shall attempt to make this paper as self-contained as practical, but
to read it in detail, the reader should have at least a general familiarity
with DAJ.  Since DAJ contains some mathematical material which seems to me
irrelevant and distracting, it may save some time to skim the present work
first, then read DAJ, then return to the present work.  

Also, DAJ, though nominally self-contained, implicitly requires some 
familiarity with the ideas of ``weak values'' of quantum observables.
The seminal paper introducing ``weak values'' is \cite{AAV}, but this contains
some questionable mathematics and is set in an infinite-dimensional context
which (in the light of subsequent simplifications) 
is unnecessarily complicated.  A leisurely presentation of my view of 
the main ideas
can be found in \cite{parrott1} and a more compact presentation in the style
of research papers in \cite{parrott2}. 

DAJ employs a common physics notation which seems to me overly complicated,
hard to typeset, and not especially well-suited to the presentation
of fundamental ideas.   
However, since the present work is largely an analysis of DAJ, I hesitate
to entirely abandon their notation.  After some thought, I decided 
to use my preferred notation (the notation of \cite{parrott1} and 
\cite{parrott2}) for the preliminary and subsequent discussions, but to switch
to theirs for the detailed analysis of their paper. 
Equation numbers (1) through (9) will be reserved for equations in
DAJ (which have the same number there); new equations in the present work
will be given numbers (100) and above. 

The next section reprints the ``Notation'' section of \cite{parrott2}.
A following ``Preliminary Summary'' section briefly sketches my view
of traditional ``weak measurement'' theory as it relates to DAJ. 
The analysis of DAJ 
starts in Section \ref{sec: DAJarg}, and readers already familiar
with DAJ may prefer to start there. 

Finally, perhaps an apology concerning typography is in order.  I tried
to use the notation of DAJ in analyzing it, but had some problems duplicating
their symbols exactly using laTeX.  
They use boldface for operators, but when I specified
boldface in laTeX ``math mode'', the symbols
came out in the ``text'' font rather than ``math font''; 
e.g., $\dajE$ instead of
{\boldmath $\hat{E}$}.  I wrote the paper that way, and only later found out
how to exactly reproduce most of their symbols.  I have not attempted to
correct this because the existing symbols seem close enough to theirs and
changes sometimes have unanticipated effects.  Anyone who has worked 
with TeX/laTeX should understand.   
\section{``Introduction and notation'' 
reprinted from \protect{\cite{parrott2}}} 
\label{sec: intro}
We assume that the reader is familiar with the concept of ``weak value''
of a quantum observable.  This concept was introduced in the seminal paper
\cite{AAV} of Aharonov, Albert, and Vaidman, called  ``AAV'' below.  It 
will be briefly reviewed below, and
a much fuller presentation intended for those  unfamiliar with weak values 
can be found in \cite{parrott1}. 

	We attempt to stay as close as possible to traditional physics notation,
	reverting to notation more common in mathematics only when it seems 
	less ambiguous or complicated. 
	Our mathematical formulation of quantum mechanics 
generally follows that of 
	Chapter 2 of the  book of Nielsen and Chuang \cite{N/C}, 
with differences in notation noted below.  

	The inner product of vectors $v, w$ in a complex Hilbert space $H$  will be
	denoted $\lb v, w \rb$, with the physics convention that this be linear in
	the second variable $w$, and conjugate-linear in the first variable $v$.
	The norm of a vector $v$ will be denoted as 
	$ |v| := \lb v,v \rb ^{1/2}$.  

Technically, a (pure) ``state'' of a quantum system with Hilbert space $H$
is an equivalence class of nonzero 
vectors in $H$, where vectors $v, w \in H$ are
equivalent if and only if $w = \alpha v$ for some nonzero constant $\alpha$.
However, we informally refer to vectors in $H$ as
``states'', or ``pure states'', when we need to distinguish 
between pure states and ``mixed states'' (see below).  
A state $v$ is said to be {\em normalized} if $|v| = 1$.   
We do not assume that states are necessarily normalized.

	The projector to a subspace $E$ will  be denoted $P_E$, in 
	place of the common but unnecessarily complicated 
	physics notation $\sum_i | e_i  \rb \lb e_i |$, where $ \{e_i\}$ is an 
	orthonormal basis for $E$.  
	When $E$ is the entire Hilbert space of states, $P_E$ is called the 
	{\em identity operator} and denoted $I := P_E$.   
	When $E$ is the one-dimensional subspace
	spanned by a vector $w$, we may write $P_w$ for $P_E$.  
	When  $|w| = 1$,  $P_w v = \lb w, v \rb w$, but the reader should
	keep in mind that   under our convention, $ P_w = P_{w/|w|}$, so this 
	formula for $P_w$ only applies for $|w| = 1$. 

Mixed states are represented
by ``density matrices'' $\rho : H \rightarrow H$, which are defined as 
positive operators on $H$ of trace 1.  
A pure state $h \in H$ corresponds to the density
matrix $P_h$. 

We shall be dealing with a quantum system $S$ in which we are primarily 
interested, which will be coupled to a quantum ``meter system'' $M$.  
We make no notational distinction between the physical systems $S$ and $M$
and their Hilbert spaces. 

The composite system of $S$ together with $M$
is mathematically represented by the Hilbert space tensor product 
$S \tnr M$.  We assume the reader is generally familiar with the 
mathematical definition of $S \tnr M$.  The highlights of the definition
are as follows.

Some, but not all, vectors in $S \tnr M$ can be written in the form 
$s \tnr m$ with $s \in S$ and $m \in M$; these are called ``product states''.  
Typical physics notation for $s \tnr m$ might be 
$|s\rb |m\rb$ or 
$|s\rb_S |m\rb_M$. 
Every vector $v$ in 
$S \tnr M$ is a (possibly infinite) linear combination of product states
:  $v = \sum_i s_i \tnr m_i$.

If $\rho$ is a density matrix on $S \tnr M$, its partial trace with respect
to $M$, denoted $\trM \rho : S \rightarrow S$ is a density matrix on $S$.
The (mixed) state of $S$ corresponding to $\rho$ is $\trM \rho$.  
\section{Preliminary summary and intended audience for this work}
\lbl{sec: prelim}
Prospective readers will naturally want to know what will be the payoff 
for their valuable time invested in understanding this paper.  This section
attempts to address this issue in a more complete way than an abstract.  

%Readers who are mainly interested in applications of  DAJ 
%to traditional ``weak value'' theory
%may be best advised not to invest 
%time in the present work because one of
%its conclusions will be that the methods of
%DAJ and its results, though interesting, 
%unfortunately offer little in the way of new understanding
%of weak values.  It would be a disservice to potential
%readers to obscure this fact.  I sincerely hope it will not be interpreted 
%in any unpleasant way.   

The abstract of DAJ states: 
\begin{quote}
``We introduce contextual values as a generalization of the eigenvalues 
of an observable that takes into account both the system observable 
and a general measurement procedure. 
This technique leads to a natural definition 
of a general conditioned average that converges uniquely 
to the quantum weak value in the minimal disturbance limit. 
As such, we address the controversy in the literature 
regarding the theoretical consistency of the quantum weak value 
by providing a more general theoretical framework 
and giving several examples of how that framework relates to 
existing experimental and theoretical results.''   
\end{quote}
I was intrigued by this abstract and assume others may be also.
Such potential readers, should be aware that questionable mathematics 
renders some conclusions of DAJ unproved, and some
are arguably false.  In particular, the present work
presents a counterexample to their claim that
their ``general conditioned average''  ``converges uniquely 
to the quantum weak value in the minimal disturbance limit''
(using a definition that seems appropriate for their
otherwise undefined  ``minimal disturbance limit'').  

 However, I still find their ideas intriguing,
and I wonder if a reformulation might yet prove useful.  Those who 
enjoy playing with ideas may find some interest both in DAJ and in 
the present work.

Following is a thumbnail sketch of my interpretation of 
the arguments of DAJ and my conclusions about them.  They will probably
not be meaningful to someone unfamiliar with weak values.
It is hoped that they may give the reader a sense of what is to come,
should he decide to invest the time in a careful reading.

First we review some key ideas of weak measurement, suppressing
technical details.  We are given the Hilbert space $S$ for
a quantum system of interest (also denoted $S$), 
an arbitrary number of copies
of some  pure state $s$ of $S$, and an observable
(Hermitian operator) $A$ on $S$.  Our object is to measure the expectation
$\lb s, As \rb$ of $A$ in the state $s$, 
denoted $\lb s, As \rb$ while negligibly changing 
the copies of $s$.   

In the ``weak measurement'' literature,
this  is usually accomplished
 by coupling $S$ to a ``meter system'' $M$, obtaining a
composite system mathematically described as the tensor product $S \tnr M$
of $S$ with $M$, and measuring a new observable $B$ on $M$ (actually,
measuring $I \tnr B$ on $S \tnr M$) in a slightly entangled normalized 
(pure) state 
$e = e(s, \epsilon)$ of $S \tnr M$, depending on $s$ and a small real
parameter $\epsilon$.%  
\footnote{DAJ calls this parameter $g$, a fact which I had not noticed
while writing the above.  I have decided not to change the present 
$\epsilon$ to $g$ because that risks introducing errors in case 
some substitutions are overlooked.
}
Here ``slightly entangled'' means that 
$\lim_{\epsilon \goesto 0} \trM P_{e(s, \epsilon)} = P_s$, where 
$P_v$ denotes the projector to the one-dimensional subspace spanned by 
the vector $v$.

It is possible to choose $B$ and $e(s, \epsilon)$ 
so that 
\beq
\lbl{eq95}
\lim_{\epsilon \goesto 0} \trM P_{e(s, \epsilon)} = P_s
\eeq
and 
\beq
\lbl{eq97}
\lb s, As \rb = \lim_{\epsilon \goesto 0} \lb e(s, \epsilon), 
(I \tnr (B/ \epsilon ) e(s, \epsilon) \rb   
\q.
\eeq

Measuring $B$  or $B/\epsilon$ 
is essentially the same as measuring the spectral projector-valued 
measure $\{Q_j\}^n_{j=1}$ associated with $B$.%  
\footnote{Here we are taking some liberties
for ease of exposition:  Strictly speaking, 
we are measuring $I \tnr B$ and $\{I \tnr Q_j \}$rather 
than $B$ and $\{Q_j\}$, 
and we are implicitly assuming a finite-dimensional context so
that the spectral measure of $B$ is finite.
}  
The result of the measurement  
of the spectral projectors 
is a new (unnormalized) state $(I \tnr Q_j) e $ with probability
$|(I\tnr Q_j) e|^2$.  The (generally mixed ) state of $S$ 
is then obtained by tracing over $M$:
$$
\mbox{(unnormalized) state of $S$} = \trM [P_{(I\tnr Q_j)e} ] 
\q. 
$$

In practice, the map $s \mapsto e(s, \epsilon)$ is usually 
obtained by applying
an $\epsilon$-dependent isometry%
\footnote{An isometry $V$ from a Hilbert space $H$ to a Hilbert space
$K$ is an operator which preserves the inner product:  
$ \lb Vh_1, Vh_2 \rb = \lb h_1, h_2 \rb$ for all $h_1, h_2 \in H$.  The
difference between an isometry and a unitary operator is that the isometry
is not assumed to map $H$ {\em onto} $K$, i.e., $VH$ need not be all 
of $K$.  
} 
 $U(\epsilon): S \rightarrow S \tnr M$ to $s$:  
$e(s, \epsilon) := U(\epsilon) s$.  Assuming this, for fixed $\epsilon$, 
the map $P_s \mapsto P_{e(s, \epsilon)}$ is easily seen to extend to
a completely positive map from (mixed) states of $S$ to states of $S \tnr M$.
Since  tracing is also completely positive,  
the composite map  
$$
P_s \mapsto P_{e(s, \epsilon)} \mapsto  
 \trM P_{(I\tnr Q_j)e(s, \epsilon)} 
$$
extends to a completely positive map on states of $S$.  
Then 
Stinespring's \cite{stinespring} or Choi's \cite{choi} theorem implies that, 
for each $j$ there exist operators
$M_{j,1}, M_{j,2}, \ldots M_{j, k_j}$ on $S$, $ j=1, \ldots , m$, such 
that for all $s \in S$,  
\beq
\lbl{eq100}
\trM [\, P_{I \tnr Q_j} P_{e(s, \epsilon)}  ] = 
\sum_{i=1}^{k_j} M_{j,i} P_s M^\dag_{j,i}
\q.
\eeq
\setcounter{myeq}{\value{equation}}
For simplicity of notation, I prefer to leave states unnormalized unless 
the context requires normalization.  After normalization, the operators
$M_{j,i}$ will additionally satisfy
\beq
\lbl{eq102}
\sum_{j=1}^m \sum_{i=1}^{k_j} M^\dag_{j,i} M_{j,i} = I
\q,
\eeq
where $I$ denotes the identity operator.
  
It is common in the physics literature
(though mathematically incorrect) 
to pretend that the sum in equation \re{eq100} only has one term,
in which case one may rename the $M_{j,i}$ as $M_j$ and (assuming \re{eq102})
call them ``measurement operators''.%
\footnote{This may have
something to do with the fact that the 
influential and generally excellent 2000 book \cite{N/C} of Nielsen and 
Chuang does this without mention, referring to the (renamed) $M_j$
as ``measurement operators''.  Strangely, a 1997 paper coauthored by Nielsen 
\cite{nielsen/caves} (``Quantum measurements'' section, equation (3.1))
clearly indicates that he was perfectly
aware that \re{eq100} is the correct formulation.
}
DAJ assumes this simplification without mention, and for notational 
simplicity, we shall also.  No essential change in most (but not all) 
of the arguments
of DAJ would be necessary if the more general \re{eq100} were used.%
\footnote{DAJ mainly uses only the positive-operator-valued measure (POVM)
 $E_{j,i} :=  M^\dag_{j,i} M_{j,i}$ associated with the $M^\dag_{j,i}$, 
which for many of their purposes can be replaced by a new singly-index POVM
$E_j := \sum_i M^\dag_{j,i} M_{j,i}$.  For  issues regarding 
``weak'' measurement (which they do not treat in a precise way), it would
probably be necessary to use the more
general doubly-indexed form.
}

Assuming this simpification, with the projector-valued measure $\{Q_j\}$
for $B$ is associated the positive-operator-valued measure (henceforth
called a POVM) $\{M^\dag_j M_j\}$.  DAJ renames this POVM as $\{E_j\}$,
with $E_j := M^\dag_j M_j$.  

Then when $S$ is in a pure state $\rho = P_s$, the probability $p_j$ that
the $j$'th measurement outcome occurs is

$$
p_j = \tr [ (I\tnr Q_j) P_{e(s, \epsilon)} ] =
\trS \trM [ (I\tnr Q_j) P_{e(s, \epsilon)} ] 
\approx \trS [  E_j P_s] 
\q,
$$ 
with the approximation becoming exact as $\epsilon \goesto 0$.  The 
corresponding equation for $S$ in a mixed state $\rho$ in the limit
$\epsilon \goesto 0$ is
$$
p_j = \tr [E_j \rho ]
\q.
$$

DAJ emphasizes that what is mainly experimentally 
accessible are the probabilities $p_j$, and these may be obtained either
from the projector-valued measure associated with the $Q_j$ or the  
POVM $\{E_j\}$.  The numerical values $\alpha_j$ which are averaged 
with the $p_j$ to obtain the expectation of $B/\epsilon$ 
are less fundamental.  

To put the matter more picturesquely, imagine that measuring $B/\epsilon$ 
entails 
recording the position of a meter pointer on a scale (with only finitely many
pointer positions possible due to our assumption of finite
dimensionality).  We can change the scale on the meter if we want, effectively
obtaining a new meter.  If the original meter did not always average
to the expectation $\tr [A\rho]$ of $A$ in the state $\rho$, we could ask
if it were possible to change the scale so that it did average to
 $\tr [ A \rho]$ for all states $\rho$.  
The new scale values (if they exist) are called
``contextual values'' by DAJ.  

Since reading the meter in $M$ is the same as assigning real values to  
the outcomes $j$ of the POVM $\{E_j\}$  in $S$ we could dispense with the
$M$ meter and work entirely in $S$, in terms of a given measurement
operators $\{ M_j \}$.  
DAJ explores the possibilities of such an approach.

Note that in this formulation, it would not be necessary 
to assume that the measurement is ``weak'' (i.e., that it did not 
significantly alter the state of $S$), and through their equation (6),
DAJ does not make this assumption.  However, outside the context of 
weak measurement, it seems unclear what would
be the advantage of using general measurement operators 
in place of normal ``strong'' measurements obtained from the spectral 
measure of $A$, and DAJ does not discuss this question.  

Eventually, DAJ adds a ``weakness'' assumption to make contact with 
traditional ``weak measurement'' theory.  Unfortunately, their particular
notion of ``weakness'' (which they call ``minimal disturbance'') 
is never clearly defined in their paper, forcing 
the reader to guess at their definitions.
We shall point out that under some reasonable guesses, their conclusions
would be false.

Some believe the basic idea of traditional ``weak measurement'' theory
to be of questionable utility 
(cf.\  the ``Remarks'' section of \cite{parrott2}).  However, those 
who do think it potentially useful may find attractive DAJ's idea of doing it
entirely within the original system $S$ of interest, without reference
to an external ``meter system''.  
Those interested in extending the work of DAJ may 
be interested in learning of its gaps from the present work.  
\section{The main argument of DAJ}
\label{sec: DAJarg}
We shall use the notation of DAJ and reproduce
their equations as written there.  We also use their equation numbers,
which range from (1) to (9).  Thus this section will probably only be
meaningful to readers who have some familiarity with DAJ and have it
at hand.  In addition, we assume the reader has at least skimmed
the ``Preliminary summary'' section above.

DAJ  attempts to implement weak measurements via ``measurement operators''
${\cal M} = \{\dajMj\}$,
seemingly using the definitions of the book \cite{N/C} of Nielsen and
Chuang (though this reference is not specifically cited).  
It is known (\cite{N/C}, pp. 94-95) that 
such measurement operators can always be obtained by coupling the 
system $S$ of interest to an ancillary ``meter system'' $M$ obtaining
a new system $S \tnr M$), performing a projective measurement in $S \tnr M$,
and then tracing out $M$ to obtain measurement operators on $S$, as
described in the ``Preliminary summary'' section.  Thus I would not expect
to  obtain anything essentially new from the formalism of DAJ, 
though it might turn
out to be more insightful or convenient than traditional 
``weak measurement'' approaches.

The discussion of DAJ on their page 2 attempts to obtain the expectation
$\lb {\cal A} \rb$ in a given state $\rho$ 
of an observable ${\cal A}$ (corresponding to operator  $\dajA$) 
from the above measurement operators and the 
associated positive-operator-valued measure (abbreviated POVM) 
$\dajEj := \dajMj^\dag \dajMj$, which leads to their equation (2):    
\setcounter{saveeq}{\value{equation}}
\setcounter{equation}{1}
\beq 
\lbl{eq2}
\lb {\cal A} \rb = \sum_j \alpha_j P_i = \sum_j \alpha_j 
\tr [\dajEj \hat{\bf \rho} ] 
\eeq 
for real numbers $\alpha_1, \alpha_2, \ldots, \alpha_N$ called 
{\em contextual values} and abbreviated CV.
(The $P_i$ in the middle term represent probabilities and should be $P_j$.)

Denoting the eigenvalues of the operator $\dajA$ by $a_1, a_2, \ldots, a_m$,
and the corresponding spectral projectors by $\dajPij$, 
equation (2) leads to their equation (4):
\setcounter{equation}{3}
\beq
\lbl{eq4}
\dajA = \sum_j \alpha_j \dajEj = \sum_k a_k \dajPik ,
\eeq
which in turn leads to a system of $m$ equations in $N$ unknowns.  
I initially interpreted the subsequent discusssion as claiming
that (4) always has a solution (obtained from the Moore-Penrose
pseudoinverse) when $N \geq m$, and 
I still think the paper does indeed give this impression. 
However, the authors have informed me (and authorized me to say) 
that instead, (4) was
intended as a {\em hypothesis} for the subsequent discussion, and that 
the sentence just below (4) was intended to imply this.%  
\footnote{A simple example for which (4) has no solution 
is $\dajEj := I/N$ with $A$ not a multiple of the identity $I$ (and using
a trace normalized to $\tr I = 1$).  
}

They then propose (without giving any substantial reason) that 
``the physically sensible
choice of CV [contextual values] is the least redundant set uniquely related
to the eigenvalues through the Moore-Penrose pseudoinverse''.
(``Least redundant set'' is not defined.)
A long paragraph then presents the Moore-Penrose pseudoinverse 
in a complicated way which obscures its essential simplicity.%
\footnote{In fairness to the authors, it should be noted 
that such complicated presentations 
are common in the physics literature,  
so they are hardly alone.  Their presentation uses the singular value
decomposition which is presented in a similarly complicated way in
the influential and mostly excellent book \cite{N/C} of Nielsen and 
Chuang.  And the presentation of the Moore-Penrose pseudoinverse 
in Wikipedia as of this writing (Jan. 23, 2011)
is similarly and unnecessarily complicated.
\s
An appendix discusses the Moore-Penrose pseudoinverse in the simple
way that most mathematicians probably view it.  I took the time to write
it in the hope that it may be helpful to those unfamiliar with this and
similar concepts.
} 

{\em If} there is a solution 
$\vec{\alpha} = (\alpha_1, \ldots, \alpha_N)$ to (4) 
(as they assume), then indeed $\vec{\alpha}_0 := \dajF^+ \vec{a}$ 
is also a solution,
where $\dajF^+$ denotes the Moore-Penrose pseudoinverse.  
But it is obscure to me why $\dajF^+ \vec{a}$ 
should be considered as better or more 
fundamental than any other solution.  It seems to me that the whole discussion
of the Moore-Penrose pseudoinverse is distracting and essentially irrelevant
to the rest of the paper.

Before continuing with the arguments of DAJ, let us  put into perspective
some ideas which have been presented previously, both in this and the
``Preliminary summary'' sections.  DAJ assumes that (4) always has a 
solution $\vec{\alpha}$.  
(Whether it is given by the Moore-Penrose pseudoinverse 
is irrelevant.)  ``Weak measurement'' theory gives something analogous
for approximate solutions to (4). 

More precisely, weak measurement theory gives measurement operators 
$\dajMj = \dajMj (g)$ and contextual values $\alpha_j = \alpha_j (g)$ 
which depend on a small ``weak measurement'' parameter $g$ and which 
approximately satisfy (4), the approximation becoming exact in the limit 
$g \goesto 0$.  
In addition, the disturbance of a given  state $\dajrho$
by the measurement goes to 0 as $g \goesto 0$. 

Therefore, in anticipation of applications of contextual value theory 
to weak measurement theory, we ought to have in mind measurement operators
$\dajMj (g)$ and contextual values $\alpha_j (g)$ which depend on a small
parameter $g$.  DAJ does introduce this generalization later.  

Before continuing the general discussion, it will
be enlightening to touch on the paper's example entitled ``Photon polarization''
which interprets an experiment of Pryde, {\em et al.}, \cite{pryde}
in the context of CV's.  Here the measurement operators and associated POVM
can be obtained in the exactly the way described in the ``Preliminary 
summary'' section, though DAJ does not indicate how they were obtained.  
And, the CV's and POVM do depend on $g$.  There
are just two CV's denoted $\alpha_+ (g), \alpha_- (g)$, and it turns out
that $\alpha_{\pm} (g) = \pm 1/g$.  
This example teaches us that in general, we can expect 
$\lim_{g \goesto 0} \alpha_j (g) = \infty$, and we have to expect
that arguments which would work for bounded $\alpha_j (g)$ may fail.

Now we continue with the paper's argument.  
It renames the previous measurement
 operators, $\calM = \{ \dajMj\}$ as   
 $\calM^{(1)} = \{ \hat{\bf M}^{(1)}_j\}$ 
and considers a second set 
of ``postselection'' measurement operators which it denotes   
 $\calM^{(2)} = \{ \hat{\bf M}^{(2)}_f \}$.

Though it does not say so, this second set of measurement operators
is presumably $\{P_f, (I-P_f) \}$, where $P_f$ is the projector on 
a postselected final state $f \in S$.%
\footnote{  
Actually, DAJ appears to use $f$ as an index rather than as a postselection
vector, but our change of notation seems likely to cause
less confusion than renaming the postselection vector something like
$h$, in which case DAJ's ${\bf E}^{(2)}_f$ would equal $P_h$ for some
choice of index $f$.   
} 
Anyway, this is the case of interest
in applying DAJ's theory of contextual values to traditional weak measurement
theory, and for simplicity I shall assume this case below. 

It considers first making a measurement described by $\calM^{(1)}$, then
a subsequent measurement described by  $\calM^{(2)}$.  The composite
measurement is described by measurement operators which DAJ denotes 
$\calM^{(1,2)} =  \{\dajM^{(2)}_j \dajM^{(1)}_f \}$.  The associated 
POVM is  $\{ \hat{\bf E}^{(1,2)}_{jf} = 
\hat{\bf M}^{(1)\dag}_j 
\hat{\bf M}^{(2)\dag}_f 
\hat{\bf M}^{(2)}_f 
\hat{\bf M}^{(1)}_j 
\}
$
Before continuing, note that in the traditional context
of weak measurements, these composite measurement operators correspond
to making an initial measurement in the {\em meter system} $M$ and then
postselecting in $S$.   

DAJ then states what it characterizes as its ``main result'', 
its equation (6) giving the ``conditioned [on successful postselection]
average of [$\dajA$]'':%
%
%\footnote{
%In DAJ, the subscript $f$ is placed to the left rather than right of 
%$\lb {\cal A} \rb$, 
%but I do not know how to accomplish that in my  older
%version of LaTeX
%and do not want to spend time  programming a new symbol.
%}
\setcounter{equation}{5}
\beq
\lbl{eq6}
_f \lb {\cal A} \rb = \sum_j \alpha^{(1)}_j P_{j|f}
= 
\frac{\sum_j \alpha^{(1)}_j \tr [ 
\hat{\bf E}^{(1,2)}_{jf} \dajrho]
}
{\sum_j \tr [ 
\hat{\bf E}^{(1,2)}_{jf} \dajrho]
}.
\eeq 
In the context of traditional weak measurement theory, equation (6)
gives the average value of the {\em meter measurement} (whether ``weak'' 
or not) conditional
on successful postselection.   Since the measurement operators
depend on what kind of meter measurement is performed, it may be
potentially misleading to characterize (6) 
as a conditional average of $\cal A$.
Given the measurement operators, the right side of (6) 
surely is the ``conditioned average'' of meaurements derived from these 
operators,  but different measurement operators satisfying (4) may
conceivably give different conditional averages.  DAJ does make this clear
in the surrounding text,
but it seems to me that denoting any of these conditional averages by 
a symbol like $_f \lb {\cal A} \rb$ which involves only ${\cal A}$ and 
$f$ and not the measurement operators invites reader misinterpretation.     
For instance, in the ``Photon
polarization'' example of Pryde, {\em et al.} \cite{pryde} mentioned above,
the measurement operators arose from the particular meter measurement chosen,
but had Pryde, {\em et al.}, chosen a different meter measurement,
a different weak value (``conditioned average'') might have been obtained.  

DAJ claims that (6) implies (7), under a hypothesis that 
``the state [$\dajrho$]
is minimally disturbed'' which they never precisely define: 
\setcounter{equation}{6}
\beq
\lbl{eq7}
A_w = \frac{\tr [ \hat{\bf E}^{(2)}_f \{ \dajA, \dajrho \} ]}
{2 \tr [\hat{\bf E}^{(2)}_f \dajrho ]} \q, 
\eeq
\setcounter{equation}{\value{saveeq}} 
where $\{ \cdot , \cdot \}$ denotes the anticommutator,
i.e., $\{B,C\} := BC + CB$.  
A little algebraic manipulation shows that 
this is a generalization to mixed states $\dajrho$ 
of the traditional weak value \re{eq1} for  pure states. 

I cannot follow DAJ's argument leading from (6) to (7), and have been
unable to guess a precise meaning for the ``minimally disturbed'' hypothesis
which could enable their argument.  This is the possibly fundamental 
gap in their argument previously mentioned.   
A later section
will present a counterexample  showing that (6) does not imply (7) for
a generalization of their ``Photon polarization'' example, using what
seems to me a reasonable (and nearly unique reasonable) 
definition of ``minimal disturbance''.

Recall the earlier remark that in a ``weak measurement'' setting,
the measurement operators $M_j = M_j (g)$ and contextual values
$\alpha_j = \alpha_j (g)$ must implicitly depend on a small real parameter
$g$ corresponding to the ``weakness'' of the measurement. 
Up to equation (6), this parameter has been unrecognized in DAJ (because
their setup does not require that measurements be ``weak'').
However the paragraph of DAJ following (6), labeled ``{\em Weak values}'', 
does explicitly introduce the parameter $g$ and sketches an argument intended
to show that (6) implies (7) under an additional ``minimal disturbance''
hypothesis which they never precisely formulate. 
The following intends to summarize and analyze their argument 
which I cannot follow in detail. 

They write the original measurement operators as  
$\hat{\bf M}^{(1)}_j (g) = \hat{\bf M}_j (g) = 
\dajUj (g) \dajEj^{1/2} (g)$, where 
$\dajUj(g) $ 
is unitary.  
This polar decomposition is surely possible.  Then they claim that 
Stone's theorem can be applied to write 
$\dajUj(g) = \exp (ig \dajGj)$ for Hermitian operators $\dajGj$.
But Stone's theorem only applies to unitary {\em groups}, i.e., it assumes
that $\dajUj(g_1 + g_2) = \dajUj(g_1) \dajUj(g_2)$, which is surely 
not true in the general context of DAJ.  For example, if it happened to 
be true for some particular measurement operators, it could be made false
for others by nonlinearly rescaling the parameter $g$.

Even if it were true, I don't understand its relevance to the problem
at hand.  They go on to state that ``$\ldots$ if $\forall j$,
$[\dajGj, \rho] = 0$, so the state is minimally disturbed, then
$\ldots$ the generalized WV [their ref. [13]] is uniquely defined as 
\setcounter{equation}{6} 
\beq 
A_w = 
\frac{\tr [ \hat{\bf E}^{(2)}_f \{ \dajA, \dajrho \} ]}
{2 \tr [\hat{\bf E}^{(2)}_f \dajrho ]} \q, 
\eeq 
where $\{ \cdot , \cdot \}$ denotes the anticommutator.''

But shouldn't ``minimal disturbance'' require that $\dajrho$ (nearly) commute
with the $\dajEj$ {\em as well as} with the $\dajUj$, or equivalently,
with the measurement operators?  
%(This is discussed in more detail below.) 
If so, it is hard to understand 
the utility of introducing the polar decomposition.  
Even supposing that $\dajrho$ can be assumed to (nearly) commute with
the $\dajUj$, 
 do the authors additionally assume its (near) commutativity 
with the $\dajEj$?  In that case, why not equivalently assume 
(near) commutativity with the measurement operators to start? 
And even if one does have some sort of near commutativity with 
the measurement operators,
since the contextual values $\alpha^{(1)}_j (g)$ can become unbounded
for small $g$, some additional argument would seem necessary to 
obtain DAJ's  conclusion that (6) converges to (7). 

Finally, let us expand on the ``minimal disturbance'' assumption.
The most plausible way to make this precise seems to me the following.
 
If the $j$-th measurement outcome is obtained, the subsequent state
is 
$$
\frac{\dajMj (g) \dajrho \dajMj (g)^\dag 
} 
{\tr [ { \dajMj (g) \dajrho \dajMj (g)^\dag ]}
} 
\q,
$$
assuming that the denominator does not vanish.  
So, I would define ``minimal disturbance'' to a state $\dajrho$ as 
\setcounter{equation}{\value{saveeq}}
\beq
\lbl{eq115}
\lim_{g \goesto 0} 
\frac{\dajMj (g) \dajrho \dajMj (g)^\dag 
} 
{\tr [ { \dajMj (g) \dajrho \dajMj (g)^\dag ]}
} 
= \dajrho \q 
\mbox{for all $j$},
\eeq
assuming that the denominator does not vanish identically in a neighborhood
of $g = 0$.%
\footnote{
If the denominator does vanish for some (but not all) $g$ near 0, 
the limit is to be understood as taken over all $g$ for which the denominator
does not vanish.  
\s
If the denominator vanishes for all $g$ near 0, then we leave ``minimal 
disturbance to a state $\dajrho$'' undefined.  Physically, this would 
correspond to zero probability for result $j$ for small $g$, 
so that  for all practical purposes, for that state $\dajrho$, 
the result $j$ could simply be deleted from the list of possible 
measurement results.
}  

A later section gives a counterexample showing that (6) does not imply (7)
under this definition of ``minimal disturbance''.   I know of no 
other plausible definition under which DAJ's argument
that (6) implies (7) would make sense. 
\section{Hermitian measurement operators}
\lbl{sec: Hmeas}
This section examines the special case of {\em Hermitian} measurement operators 
$M_j$ under our ``minimal disturbance'' assumption \re{eq115}.  
For example, this is the case for DAJ's ``Photon polarization''
example.  
The counterexample of the next section will require the result of 
this example, and it is enlightening and not much more trouble to   
work out part of it in a more general context.

To avoid distracting degenerate cases 
(cf. the footnote to the definition of ``minimal disturbance'' of the 
last section),
it will be assumed below
that the denominator of our proposed ``minimal disturbance'' condition 
\re{eq115} never vanishes. 
For the application of the results of this section to the counterexample
of the next section, no degenerate special cases occur, 
and this assumption is unnecessary.

Recall that for simplicity we are assuming that the final measurement
is postselection to a final state $f \in S$, 
so that the only final measurement operator of interest is $P_f$.
This makes the superscripts $(1)$ and $(2)$ on the various quantities in (6) 
superfluous and allows us to write (6) in the simpler-appearing form: 
\beq
\lbl{eq120} 
_f \lb {\cal A} \rb = 
\frac{ \sum_j  \alpha_j \tr [P_f 
\dajMj \dajrho 
\hat{\bf M}^\dag_j 
]
}
{\sum_j 
\tr  [ P_f \dajMj \dajrho \hat{\bf M}^\dag_j ] 
}
\q,
\eeq
where we have used $P^2_f = P_f$ and 
the cyclic property of the trace to suppress
a $P_f$ on the right. 

Including the ``weak limit'' parameter $g$, and using the 
``minimal disturbance'' assumption \re{eq115},  
the denominator of \re{eq120} can be written 
in the limit $g \goesto 0$ as
\begin{eqnarray} 
\lbl{eq122} 
\lefteqn{
\lim_{g \goesto 0}
\sum_j \tr [P_f \dajMj (g) \dajrho \hat{\bf M}^\dag_j (g)] 
=
} 
\nonumber\\ 
&&
\lim_{g \goesto 0} \sum_j \tr [ P_f 
\left( 
\frac{ 
\dajMj (g) \dajrho  \hat{\bf M}^\dag_j (g) 
}
{\tr 
[ \dajMj (g) \dajrho  \hat{\bf M}^\dag_j (g) ]
} 
- \dajrho
\right)  
 \tr [\dajMj (g) \dajrho  \hat{\bf M}^\dag_j (g)] 
 \nonumber 
 \\ 
&& 
\q\q\q\q\q + \lim_{g \goesto 0} 
\sum_j \tr[P_f\dajrho ] \tr [\dajMj (g) \dajrho  \hat{\bf M}^\dag_j (g) ] 
\nonumber \\
&=& 
\tr [P_f\dajrho] \lim_{g \goesto 0} \tr  \sum_j  
\hat{\bf M}_j (g)^\dag \dajMj (g) \dajrho    
\nonumber \\
&=& \tr [P_f \dajrho] \q,
\end{eqnarray}
because $\sum \hat{\bf M}_j (g)^\dag \dajMj (g) = I$ and $\tr \dajrho = 1$.  
This is half the denominator 
of DAJ's  final result (7).
Note that this simplification of the denominator
did not assume Hermiticity for the measurement operators; it will
be needed later for non-Hermitian measurement operators.

Next note that part of the numerator of (6) 
can be written as: 
\beq
\lbl{eq130}
\dajMj \dajrho \hat{\bf M}^\dag_j = 
(1/2)(\, \dajMj \hat{\bf M}^\dag_j \dajrho + 
\dajrho \dajMj \hat{\bf M}^\dag_j\, ) +  
(1/2) (\, [\dajMj, \rho] \hat{\bf M}^\dag_j + 
\dajMj [\rho,  \hat{\bf M}^\dag_j] \, ) \q,
\eeq 
where the brackets denote commutators: $[B,C] := BC - CB$.
The full numerator of (6) or \re{eq120} similarly decomposes into two
terms corresponding to \re{eq130}:
\begin{eqnarray}
\lbl{eq135}
\lefteqn{
\mbox{numerator of (6)} =
}
&& \nonumber \\
&&(1/2)\tr [ P_f \sum_j \alpha_j 
\{\dajMj \hat{\bf M}^\dag_j \dajrho + \dajrho \dajMj \hat{\bf M}^\dag_j \} 
]
\nonumber \\
&&
+  
(1/2) \tr [P_f \sum_j \alpha_j 
(\  [\dajMj, \dajrho]\, \hat{\bf M}^\dag_j 
\  + \  \dajMj\, [\dajrho,  \hat{\bf M}^\dag_j]\ ) \, ]
\q.
\end{eqnarray}
Recalling that the contextual values $\alpha_j$
are assumed chosen so that $\sum_j 
 \alpha_j \hat{\bf M}^\dag_j \dajMj = \dajA$ and using the assumption 
of Hermiticity to write  
$\dajMj \hat{\bf M}^\dag_j =  
  \hat{\bf M}^\dag_j \dajMj$, the first term of \re{eq135} simplifies to
\beq
\lbl{eq137}
(1/2) \tr [P_f \{\dajA, \dajrho \}]
\q, 
\eeq 
which is half the numerator of DAJ's (7).  Thus we obtain DAJ's (7) under
the assumptions of Hermitian measurement operators and ``minimal disturbance''
\re{eq115} {\em if and only if} the second term in \re{eq135}
vanishes in the limit $g \goesto 0$.  

But ignoring that term seems problematic.  
The problem is that the contextual values $\alpha_j = \alpha_j (g)$ may 
become large as $g \goesto 0$, as occurs for DAJ's 
``Photon polarization'' example for which $\alpha_j (g) = \pm 1/g$.
So even if the ``minimal disturbance'' assumption could somehow be applied
to assure that the commutators in the second term of \re{eq135} become
small as $g \goesto 0$, 
we would still need some additional control over the rates of convergence
to assure the vanishing of the second term in the limit $g \goesto 0$.  
The lack of control over the rates of convergence seems the crucial 
difficulty in passing from (6) to (7), even for the special case of
Hermitian measurement operators.

\section{Contextual weak values are not necessarily the traditional
weak value.}
\lbl{sec: nontrad}
This section presents a counterexample showing that the difficulty 
in passing from DAJ's equation (6) to the traditional weak value \re{eq1} 
is essential if we accept the definition \re{eq115} of ``minimal disturbance''.
It is a modification of DAJ's ``Photon polarization'' 
example on their page 3, but I shall use notation easier to typeset.%
\footnote{Like many physics papers, DAJ distinguishes operators by 
{\em both} boldface and ``hats''.  Notationally distinguishing different
quantities may be helpful to readers who are skimming a paper, but surely
either boldface or hats would be enough.  For serious readers, neither
should be necessary, and I will omit them.
}

This example arises from an experiment of Pryde, et al. \cite{pryde}, 
which performs a weak measurement and obtains the traditional weak value
\re{eq1} for that measurement.  The measurement operators which DAJ
denotes $M_{\pm}$ and the contextual values (which are the eigenvalues
of the ``meter observable'' of \cite{pryde}) 
can be obtained from data given in the paper of Pryde, 
et al., as described in the ``Preliminary summary'' section around equation
\re{eq100}.  For this purpose, one starts with the slightly entangled state
called there $e(s, \epsilon)$ as given in \cite{pryde} but not mentioned
in DAJ.  For the purpose of merely checking that DAJ's (6) does imply (7), one
can start with the DAJ's measurement operators and contextual values
without considering their origin, and this is what we shall do. 

DAJ's ``Photon polarization'' example is set in a two-dimensional complex
Hilbert space $S$ with a distinguished orthonormal
basis with respect to which all matrices below will be written. 
As in DAJ, the operator $A$ for which $\lb s, As \rb$ is to be weakly measured
will have matrix with respect to this basis 
\beq
\lbl{eq120 }
A =  
\left[
\begin{array}{rr} 
1 & 0 \\
0 & -1 
\end{array}
\right]
\q.
\eeq 
Let $g$ denote a small parameter let $\lambda := \sqrt{(1+g)/2}, 
\mu := \sqrt{(1-g)/2},$ so \linebreak $\lambda^2 + \mu^2 = 1$.%
\footnote{  (DAJ 
denotes $\lambda$ as $\gamma$ and $\mu$ as  $\bar{\gamma}$, but I hesitate to do this because
the mathematical literature generally uses an overbar to denote 
complex conjugate.  Also, I accidentally substituted $\lambda$ for $\gamma$,
and don't  want to reset the type. 
}

Let $M_{\pm} $ be as given in DAJ, namely 
\beq 
\lbl{eq125}
M_+ := 
\left[
\begin{array}{rr} 
\lambda & 0 \\
0 &  \mu 
\end{array}
\right]
\q,
M_- := 
\left[
\begin{array}{rr} 
\mu & 0 \\
0 & \lambda 
\end{array}
\right]
\q.  
\eeq
The contextual values $\alpha_{\pm} = \alpha_{\pm} (g)$ given in DAJ are 
\beq 
\lbl{eq127} 
\alpha_{\pm} (g) = \pm 1/g
\q, 
\eeq
and these result in 
\beq
\lbl{eq128}
A = \alpha_+ M^\dag_+ M_+ +  
 \alpha_- M^\dag_- M_- 
= \alpha_+ M^2_+ + 
 \alpha_- M^2_-
\q \mbox{for all $g$}.
\eeq

We shall define two new measurement operators depending on $g$: 
\beq
\lbl{eq150}
M_1 (g) := U (g) M_+ (g)     
\q \mbox{and} \q M_2 (g) := M_- (g)
\q,
\eeq
with $U(g)$  unitary operators to be specified later.  Note that this 
leads to the same POVM $\{ E_1 := M^\dag_1 M_1, \  E_2 := M^\dag_2 M_2 \}$
as in DAJ, so that $\sum_j \alpha_j E_j = A$ holds in both contexts if we
define
$$
\alpha_1 (g) := \alpha_+ (g) = 1/g, \q \alpha_2 (g) := \alpha_- (g) = -1/g \q.
$$
The starting point for the counterexample will be
the fact that when $U(g) = I$ for all $g$, equation (6) of DAJ does
lead to the traditional weak value given in its equation (7).

To check this, verify that for an arbitrary mixed state 
$$\rho = 
\left[
\begin{array}{rr}
\rho_{11} & \rho_{12} \\
\rho_{21} & \rho_{22}
\end{array}
\right]
\q,
$$ 
$$
[M_+, \rho] = 
\left[
\begin{array}{cc}
0 & (\lambda - \mu) \rho_{12} \\
(\mu - \lambda) \rho_{21} & 0 
\end{array}
\right]
\q,
$$
and
$$
 [M_+, \rho] M_+  \ + \  M_+ [\rho, M_+ ] = 
\left[
\begin{array}{cc}
0  & -(\lambda - \mu)^2 \rho_{12} \\
- (\lambda - \mu)^2) \rho_{21} & 0 
\end{array} 
\right]
$$
Interchange $\lambda$ and $\mu$ to obtain
the corresponding expression with $M_+$ replaced by $M_-$, 
and the corresponding expression is the same.  
Then the expression 
$$\sum_j \alpha_j ( 
[M_j, \rho] M^\dag_j + M_j [ \rho , M^\dag_j]\, ) $$
in the second term of \re{eq135} is easily seen to vanish identically
because  $  
[M_+, \rho] M^\dag_+ + M_+ [ \rho , M^\dag_+ ] =  
[M_-, \rho] M^\dag_- + M_- [ \rho , M^\dag_- ] 
$
but $\alpha_- = - \alpha_+$.  Hence the second term of \re{eq135} vanishes
identically, and then (7) follows.

Now consider the effect of replacing the Hermitian measurement operators 
$M_+ (g) $ with new  measurement operators $M_1 (g), M_2 (g)$
with  
$$
M_2 (g) := M_- (g) \q \mbox{and $M_1 (g) := U(g) M_+ (g)$}, 
$$
with  $U(g)$ a $g$-dependent unitary operator to be determined.
We shall choose $U(g)$ so that $\lim_{g \goesto 0} U(g) $ is the identity
operator $I$ in order to satisfy our ``minimal disturbance condition
\re{eq115}.

Recall that equation \re{eq120} gives 
the expectation $_f \lb {\cal A} \rb$ of $A$ in initial mixed state 
$\rho$ and postselected to the pure state $f \in S$.  
Let $\Delta (g) $ denote the difference between the value given by this 
equation  
for the Hermitian measurement operators $M_+ (g) , M_- (g) $ and for the new
measurement operators $M_1 (g) := U (g) M_+ (g) , M_2 := M_- (g)$.
To show that (6) does not imply (7) in general, we shall show that it is
possible to choose unitary $U(g)$ and states $\rho$, $f$, such that
\beq
\lbl{eq155}
\lim_{g \goesto 0} \Delta(g) \neq 0
\q.
\eeq

Equation \re{eq122} shows that
the limit as $g \goesto 0$ of the denominator of \re{eq120} is 
$\tr [ P_f \rho ]$  
for both sets of measurement operators.  
The numerator of \re{eq120} for measurement operators $M_1, M_2$ is
\begin{eqnarray*} 
\lefteqn{
\alpha_1 \tr [P_f UM_+ \rho M_+ U^\dag] \,  + \, 
\alpha_2 \tr [P_f M_- \rho M_- ]
= 
} \\
&& 
\alpha_+ \tr [\, [P_f, U]M_+ \rho M_+ U^\dag 
+ 
\alpha_+ \tr [\,  U P_f  M_+ \rho M_+ U^\dag \, ]\, +\, 
\alpha_- \tr [P_f M_- \rho M_-] \\
&=& \alpha_+ \tr [\, [P_f, U] M_+ \rho M_+ U^\dag \,]  + 
(\alpha_+ \tr [P_f M_+ \rho M_+] +  \alpha_- \tr [P_f M_- \rho M_- ])
.
\end{eqnarray*}
The term in parentheses is
the numerator of \re{eq115} for measurement operators $M_+, M_-$.
%the difference
%plus the additional term 
%$\alpha_+ \tr [\, [P_f, U]M_+ \rho M_+ U^\dag $ .

We shall choose $U(g) := \exp (igH)$ with $H$ a Hermitian operator
which is not a multiple of the identity.  
Then the difference of the limits of the numerators of \re{eq120} for
the two sets of measurement operators is  
\begin{eqnarray}
\lbl{eq160}
\lefteqn{
\lim_{g \goesto 0} 
\tr [\alpha_+ (g) \tr [\, P_f, U(g) ]M_+ (g) \rho M_+ (g) U^\dag (g) \, ] 
= } \\
&& 
\lim_{g \goesto 0}  
\tr [(1/g) \tr [\, [P_f, U(g) ]M_+ (g) \rho M_+ (g) U^\dag (g) \, ]\nonumber \\
&=&  
\tr [\, [P_f,iH] \, \rho ] 
\q,
\end{eqnarray}
where the limit of the commutator is given by: 
\begin{eqnarray*}
\lim_{g \goesto 0} (1/g) [P_f, U(g)] &=& 
\lim_{g \goesto 0}  [P_f, (U(g) - I)/g] \\
&=& [P_f, dU(g)/dg|_{g=0} ] \\
&=&  [P_f, iH]
\q.
\end{eqnarray*}
We have shown that 
\beq
\lbl{eq170}
\lim_{g \goesto 0} \Delta(g) =  \frac{\tr [\,[P_f,iH]\,]\rho ]}
{\tr[ P_f \rho ]}
\q. 
\eeq
It is well-known and elementary to show that the only operators $H$ 
which can commute with all one-dimensional projectors $P_f$ are multiples
of the identity operator, so there exists some $f \in S$ such that 
$[P_f, iH] \neq 0$.  And then there exists some mixed state $\rho$, 
which can be chosen of the form $\rho = P_h$, such that 
$\tr [ [P_f, iH] P_h ] = \lb h, [P_f, H] h \rb/(2 |h|^2) \neq 0$.%
\footnote{The factor $1/2$ arises because we are using a trace normalized
to $\tr I = 1$.
}
The last fact follows because it is also well-known that the only operator
$B$ which can satisfy $\lb h, Bh \rb$ for all $h$ is $ B = 0$; for present
purposes we only need this for the Hermitian operator $B = [P_f,iH]$ for which
it is obvious from the spectral theorem. 
\section{Summary and conclusions}
\lbl{sec: Summary} 
``Measurement operators'' 
on a quantum system $S$ can always be implemented by  
passing to a tensor product $S \tnr M$ of $S$ with a ficticious 
ancillary quantum system $M$, performing a projective measurement
in $S \tnr M$
with orthogonal projectors of the form $\{I \tnr Q_j \}$,
and then tracing out $M$ to obtain measurement
operators on $S$ (\cite{N/C}, pp. 94-95).   By ``passing to $S \tnr M$''
we mean identifying $S$ with a subspace of $S \tnr M$ via an isometry
$V: S \rightarrow S \tnr M$.   

This gives a way to translate
the language of ``measurement operators'' 
on $S$ into the traditional language of ``weak measurement'' theory.
The projective measurement in $S \tnr M$ corresponds to a 
``meter measurement''  in $M$, where $\{ I \tnr Q_j \}$ are the spectral 
projectors for a ``meter observable'' $I \tnr B$.  
(We speak of ``weak measurement theory'' because we need some name for it,
but the translation just mentioned does not require ``weakness''
of measurements deriving from measurement operators, and we do not 
necessarily assume it below.)

For the inverse translation, one needs to replace the language of
``measurement operators'' as expounded in \cite{N/C} by something more
general (cf.\ equation \re{eq100}).  Thus the language of weak measurement
seems more general than the language of measurement operators.%
\footnote{
DAJ's claim that ``the WV [weak value] can be subsumed as a special case
in the CV [contextual value] formulation'' seems to me overreaching.  It seems
difficult to pinpoint a precise logical relation between ``contextual value'' 
and ``weak value'' theory due to lack of systematic exposition 
of both in the literature.  However, intuitively, I regard 
weak value theory as more general. 
}

DAJ attempts to do weak measurement theory entirely within $S$, without
introducing a ``meter system'' $M$, replacing it by assuming as given a set
of measurement operators.  
DAJ's equation (6) formulates 
a  definition of ``conditioned average'' of  
an observable $A$ on $S$ in terms of measurement operators, which
it denotes by notation similar to $_f \lb A \rb$.
Assuming DAJ's equation (4), the unconditioned
average obtained from these measurement operators is 
the same as the average of $A$ that would be obtained  in the usual way from  
the spectral projector-valued measure for $A$, but even assuming (4), 
the corresponding
``conditioned averages'' need not be the same (as DAJ clearly recognizes).  
The analog of DAJ's (6) 
in weak measurement theory would be the expectation of the {\em meter measurement}
conditional on successful postselection in $S$.  

Much of the traditional
weak measurement literature blurs the distinction between the latter
and what one might call the ``conditional expectation of $A$ given 
successful postselection''.  I think it worth emphasizing that DAJ 
does not make this mistake. Though it does use the potentially misleading 
notation $_f \lb A \rb$ which suppresses the measurement operators from the 
``conditioned average'' obtained from them, it explicitly points out
that $_f \lb A \rb$ does depend on the measurement operators, and in 
general cannot
be obtained from $A$ (and the initial and final states of $S$) alone.    

Next DAJ claims that in some ``minimal disturbance'' limit 
which it does not precisely define, the dependence of  its 
``conditioned average'' $_f \lb A \rb$ on the measurement operators 
washes out, and the limit is the traditional ``weak value'' (7) 
which depends only on $A$ and the initial and final states in $S$. 

However, I think that its conclusion that $_f \lb A \rb$ necessarily 
converges to the traditional weak value \re{eq1} 
in some ``minimal disturbance limit''
is still unproved and probably false under definitions 
of ``minimal disturbance limit'' which reflect the normal meanings
of the English words.  The present work proposes a definition
of ``minimal disturbance limit'' which seems physically compelling and 
under which $_f \lb A \rb$ can converge to something other than the traditional
weak value \re{eq1} in this limit. 
\section{Afterword from Version 3}
Since writing the above, I've discovered some misunderstandings 
which I need to correct.  Nothing is wrong in the preceding, but  
I know more now than when I wrote it.  Revising the entire manuscript 
would not be worth the effort, so instead I am adding this Afterword.

I acquired this additional insight in the following way.   I urged the authors
of DAJ to submit an Erratum to Physics Review Letters (PRL) 
correcting the many errors and misleading obscurities in the original.  
I was particularly
concerned that an attempted proof they had sent me that (6) implies (7) 
for positive measurement operators was incorrect.
After they made clear that they had no intention of doing so, 
I submitted a ``Comment'' paper to PRL. 
  
As is the usual policy of PRL, they first sent the ``Comment'' to the authors,
who returned a lengthy reply.  It was much clearer than the  
two previous communications which they had sent me, and in its light,
I saw that I had not properly understood their usage of the term 
``minimal disturbance limit''.  I still don't 
fully understand the sense in which they use this term, 
but I have a better idea than before. 
I cannot improve my understanding because
further inquiries to the authors have gone unanswered.

Before continuing, I want to make clear that nothing I say from here on
is authorized by the authors of DAJ.  It reflects nothing more than 
my best understanding based on what they have sent me and PRL.
Most of it comes from their reply to PRL.
A draft has been sent to the authors of DAJ.
They have not replied.

Their response to PRL claims that the term ``minimal disturbance 
measurement'' has a precise meaning in quantum measurement theory to
refer to 
``Hermitian'' (presumably, they mean ``positive'') measurement operators,
and they cite this usage in the recent book of Wiseman and Milburn 
\cite{wiseman} (which is not in the reference list of DAJ). 
The book {\em does} define
the term in this way, 
but a search of the arXiv and Internet for other instances of this or
related terms (like ``minimal disturbance limit'') and an inquiry to 
the prominent Internet ``bulletin board'' sci.physics.research have failed to
turn up a single additional reference which uses the term in a context similar
to DAJ or Wiseman/Milburn. 

They go on to say that they generalize this usage to mean that
$[\dajGj, \dajrho ] = 0$, under the assumption that 
$\dajUj (g) = \exp (ig \dajGj)$, 
and they claim that this constitutes
a precise definition in DAJ of ``minimum disturbance measurement''. The only reference
I can find to this generalization in DAJ 
is the following  sentence just before (7): 
\begin{quote}
``However, if $\forall j,  [\dajGj, \rho ] = 0,$ so the state is minimally
disturbed, then the context dependence vanishes and $\ldots$ ''. 
\end{quote}
I think it would take a mind reader to guess that this is a 
{\em definition} of ``minimally disturbed''.  The syntax would normally
be interpreted as a statement that {\em if}\ $[\dajGj, \rho] = 0$, 
{\em then} the state is minimally disturbed, where ``minimally disturbed''
had  previously been defined (either by a technical definition or implicitly
by usual English meanings of ``minimally'' and ``disturbed''). 
That is how I had interpreted it.

Note that even if we accept this as a definition of ``minimal disturbance
measurement'', we woud still need a definition of the ``minimal disturbance
{\em limit} [emphasis mine]'' which DAJ claims as a hypothesis for (7).
There is no limit in the condition   $[\dajGj, \dajrho] = 0$,
and this condition is automatically satisfied  
for positive measurement operators, so some further ``limit'' condition 
(perhaps similar to \re{eq115}) is still necessary.  Repeated inquiries
to the authors asking what condition they are using have gone unanswered. 

Note also how $\dajUj (g) = \exp (ig \dajGj)$ now appears as an 
{\em assumption} rather than as the {\em conclusion} which the previous 
text of DAJ clearly suggests:
\begin{quote}
``To find the weak limit of (6) we note that any measurement context
continuously connected to the identity operation can be decomposed
into the form ${\cal M} = \dajUj (g) \dajEj^{1/2} (g),$ where $g$ 
is a measurement strength parameter and 
$\dajUj (g) = \exp [ig \dajGj] \ldots$''
\end{quote} 
This essentially says that any real-parametrized set of unitary operators
$\dajUj (g)$ with $\dajUj (0) = I$ is a 1-parameter group of unitary 
operators (or, more sympathetically interpreted, can be reparametrized
to become a 1-parameter group).  But the unitary parts of general measurement
operators can be {\em completely arbitary}!  There is no reason that for fixed 
$j$, the different $\dajUj (g)$ should even commute, as members of a 
1-parameter group must.  

Even assuming that this was intended as an assumption, it is clearly an
extremely strong assumption which would be expected to 
hold only in unusual cases for measurement 
operators with $\dajUj (g) \neq I$.  
So it seems that the essence of DAJ's hypothesis for (7) is that the 
measurement operators $\dajMj$ be positive (which implies that they
are Hermitian).  

However, this is still an interesting hypothesis.  The reader may recall
that the ``Hermitian measurement operators'' section was unable to 
establish (7) under this hypothesis, and strengthening it to assume positive
measurement operators would not have helped.  What looked like a possibly
essential obstruction to a proof had already appeared.

In response to my earlier inquiries, DAJ had sent me an attempted proof
that (6) implies (7) assuming that $[\dajGj, \dajrho] = 0$.  There was
a serious gap in it.  I asked about the gap, and they sent me an expanded
proof to fill the gap.  But the expanded proof was definitely wrong
because there is a counterexample to one of its crucial steps, and it
looked as if the difficulty might well be essential.  

I sent them the 
counterexample and have received nothing of substance from them since.
Despite several direct inquiries,
they have neither acknowledged that the proof they sent me was incorrect
nor claimed that they still think they can prove that (6) implies (7) for 
positive measurement operators.

I want to acknowledge that 
DAJ does state that contextual weak values are not 
necessarily the traditional weak value:
\begin{quote}
``Writing the initial context in (6) to first order in $g$, we find
that as $g \goesto 0$, the weak limit generally depends explicitly 
on $\{\dajGj \}$ and $\{\alpha_j \}$ and thus will change depending
on how it is measured and how the CV are chosen (see also [11]).''
\end{quote} 
I somehow overlooked this while writing Version 2, perhaps because 
it is embedded in the vague and error-ridden paragraphs leading to (7)
of which I could make no sense.  I have changed the title and rewritten
the abstract to reflect that my Section \ref{sec: nontrad} example
that contextual weak values are not necessarily the traditional weak
value was essentially already announced in DAJ (assuming the guess that
DAJ's definition of ``weak limit'' coincides with my \re{eq115}). 

Other than that, I have left the body of Version 2 nearly intact.  A
few typos and  potentially misleading phrases have been corrected.  
\subsection{Origin of the term ``minimal disturbance measurement''}
Lacking any definition in DAJ, or even a {\em reference} to a definition, 
I had guessed that the words ``minimal disturbance limit''  were 
synonymous, or nearly so, with ``weak measurement'' as used in
works on weak measurement such as \cite{AAV}.  This turned out to be an
incorrect guess, but it seems an almost inevitable misunderstanding
given the vague way that DAJ is written.  I thought that DAJ's abstract
claimed that contextual values were always the 
``quantum [i.e., traditional] weak value'' in the limit of weak measurements. 

The motivation given in the book of Wiseman
and Milburn \cite{wiseman} for the term ``minimal disturbance measurement'' 
derives from an 
interesting paper of Banaszek \cite{banaszek}.  Using one of many reasonable 
definitions of ``closeness'' of states, he obtains the 
{\em average} over all pure states of the closeness of a postmeasurement
state to the premeasurement state, and finds that this {\em average} is 
maximized by positive measurement operators.  

But for a {\em particular} premeasurement state $\rho$ it is not necessarily
true that the postmeasurement state $\sum_j M_j \rho M^\dag_j$ is closest
to $\rho$ when $U_j = I$ in the polar decompositions $M_j = U_j H_j$ 
($U_j$ unitary, $H_j$ positive).%
\footnote{\cite{wiseman} Exercise 1.28, Section
1.4.2}
Since DAJ is dealing with {\em particular} premeasurement states 
(satisfying their ``minimal disturbance'' condition $[\dajGj, \dajrho] = 0$)
their terminology ``minimal disturbance limit'' seems not only potentially 
misleading, but actually inappropriate in their context.  
\subsection{Possible alternate definitions of ``minimal disturbance limit''}
Consider measurement operators $\dajMj = \dajMj (g)$ 
with polar decompositions $\dajMj (g) = \dajUj(g) \dajEj (g)$ with 
$\dajUj (g)$ unitary and $\dajEj (g)$ positive.  
I am now sure that DAJ's definition of ``minimal disturbance limit''
includes the condition 
$[\dajGj, \dajrho] = 0$, 
assuming
that $\dajUj (g) = \exp (i g \dajGj )$.  
This is much more restrictive than 
\beq
\lbl{eq300}
[\dajUj (g), \dajrho] = 0 \q \mbox{for all $g$}
\eeq
(which would not assume that 
$\dajUj (g) = \exp (i g \dajGj )$), or  
\beq
\lbl{eq310}
\lim_{g \goesto 0} [\dajUj (g) , \dajrho ] = 0 \q,
\eeq
which would not require that $\dajrho$ commute exactly with the unitary
parts of the measurement operators, but only in the limit $g \goesto 0$.
My original guess was that the definition of DAJ would 
include something like \re{eq310},
which seems much more natural than 
$[\dajGj, \dajrho] = 0$, 

Any positive measurement operators
(i.e., $\dajUj (g) = I$ for all $g$ and $j$) satisfy all these definitions,
so some further condition must be necessary if ``minimal disturbance limit''
is to have any meaning even remotely suggested by the words 
``minimal disturbance'' and ``limit''.
The ``Preliminary summary'' Section \ref{sec: prelim} pointed out that
in translating from the ``weak measurement'' point of view 
to DAJ's ``contextual value'' point of view, 
a projective measurement which gives  
result $j$ 
(i.e., a 
``meter measurement'') is made in a composite system $S \tnr M$ and then
$M$ is traced out to obtain a completely positive map on $S$ which sends
a premeasurement state $\dajrho$ to the (unnormalized) postmeasurement state 
\beq
\lbl{eq320} 
\sum_{i=1}^{k_j} \dajM_{j,i} \dajrho \dajM^\dag_{j,i}
\q,
\eeq
the probability of this transition being 
$$
p_j \ = \  \tr [ \sum_{i=1}^{k_j} \dajM_{j,i}\dajrho\dajM^\dag_{j,i} ]
\ = \  \tr [ \sum_{i=1}^{k_j} \dajM^\dag_{j,i}\dajM_{j,i} \dajrho ]
\q.
$$ 
Define 
\beq
\lbl{eq330}
\dajMj :=  \left[ \sum_{i=1}^{k_j} \dajM^\dag_{j,i}\dajM_{j,i} \right]^{1/2} 
\q,
\eeq
call these ``measurement operators'',  and note that they are necessarily
positive.
For the purpose of defining ``contextual values'', 
one needs only the probabilities $p_j$, which can be obtained from the 
$\dajMj$,
and so it might seem that the more general language of completely positive
maps could be replaced by the more restrictive language of ``measurement
operators'' adopted by DAJ.  If so, one could also assume that the measurement
operators are positive, and the issue of which of the above definitions
(\re{eq300}, \re{eq310} or the $[\dajGj, \dajrho ]= 0$ of DAJ) 
are used for ``minimal disturbance measurement would become moot.   

However, for the purpose of defining ``weak limit'', or ``minimal disturbance
limit'' for positive measurement operators, it may make a difference 
whether one uses the doubly indexed form \re{eq320} or traditional singly
indexed measurement operators \re{eq330}.  This is because the 
(unnormalized) postmeasurement
state when result $j$ is obtained with measurement operators $\{\dajMj \}$
is 
\beq
\lbl{eq340}
\dajMj \dajrho \dajMj^\dag
\q,
\eeq
which is not generally the same as the postmeasurement
state \re{eq320} obtained from 
the doubly indexed measurement operators.  For example, it is not 
clear that ``doubly indexed'' analogs 
of my ``minimal disturbance'' condition \re{eq115},
such as 
\beq
\lbl{eq350}
\lim_{g \goesto 0} 
\frac{\dajM_{j,i} (g) \dajrho \dajM^\dag_{j,i} (g) 
} 
{\tr [  \dajM_{j,i} (g) \dajrho \dajM^\dag_{j,i} (g) ]} 
= \dajrho, \q \mbox{for all $j, i,$ and $\dajrho$}, 
\eeq
or
\beq
\lbl{eq355}
\lim_{g \goesto 0} 
\sum_j \sum_{i=1}^{k_j} \dajM_{j,i} (g) \dajrho \dajM^\dag_{j,i} (g) 
= \dajrho, \q \mbox{for all $\dajrho$}, 
\eeq
imply \re{eq115} or conversely.

Another possible approach would be to consider the pairs $(j,i),\  
j= 1, 2, \ldots,$ $i=1, \ldots k_j$, as 
``primitive results'' which could be reindexed with a single index if 
desired.  However, since {\em all} of the results $(j,1), (j,2), 
\ldots , (j, k_j)$ in the measurement on $S$ 
correspond the the {\em single} result $j$ in the meter system $M$, 
this would require the strange assumption  that finer measurement
distinctions were possible for measurements 
in the original system $S$ than in the meter system $M$.  In that case, 
why would one bother with ``meter measurements'' in $M$ at all?  
And it seems almost a conceptual contradiction
that finer measurement results $(j,i)$ in $S$ could somehow emerge 
from the coarser results $j$ in $M$.  
It is unclear to me how this viewpoint
could be related to actual experiments. 

DAJ claims that 
\begin{quote}
``$\ldots$ [the] WV [weak value] can be subsumed as a special case in the CV [contextual
value] formalism'',
\end{quote}
but this seems to me not so obvious, and quite likely false.  A convincing
proof would surely require careful statements of assumptions and a 
careful definition of ``minimal disturbance limit''.
The converse seems more likely:  that WV theory subsumes CV theory.
\subsection{A final word}
I was trained as a mathematician and only later developed a serious
interest in physics.  I have often been appalled at the 
general unreliability of the physics literature.  Over the years,
I have learned not to waste time on papers which use 
undefined symbols or terms, and that claimed results of 
calculations can be trusted only
after one has performed them for oneself.

If I had followed that policy with DAJ, I would have avoided
a lot of work and aggravation.  But I was sucked in by the tantalizing 
but unfortunately false assumption
that ``minimal disturbance limit'' had to have a meaning something like
the ordinary meaning of the English words composing it, and the probability
that if it did, DAJ's results might be in contradiction to mine. 
I submit this to the arXiv in the hope 
that it may save other readers from similarly
wasting their time.  
\section{A possibly definitive counterexample to DAJ's (7)}
\lbl{sec: V4example}
Earlier versions of this paper noted a major gap in DAJ's passage from 
its ``general conditioned average'' (6) to what it calls the ``quantum
weak value'' (7) (which for pure states reduces to 
the traditional weak value \re{eq1}).  Versions 1 and 2 presented 
a counterexample to (7) assuming my guess \re{eq115} at the meaning
of DAJ's undefined ``minimal disturbance limit''.  But, as described
in the Afterword to Version 3 above,  my guess turned out to be wrong.

DAJ was assuming that the measurements satisfied 
a slight generalization of the definition of ``minimally disturbing
measurement'' given in Section 1.4.2 of Wiseman and Milburn's book 
\cite{wiseman}.%
\footnote{These are the same as what are called ``minimum disturbance
measurements'' in the Afterword to Version 3; the latter is the term used
in DAJ's reply to my ``Comment'' submitted to Physical Review Letters.
When I wrote the Afterword, I did not have access to Milburn and Wiseman's
book \cite{wiseman}.}   
\footnote{They do not correspond to what one might
guess given only the 
normal associations of the English phrase ``minimally disturbing
measurements''; see the Afterword to Version 3 for more information.
}
Wiseman and Milburn define a ``minimally disturbing measurement''
as one for which the measurement operators are positive.  I shall
use this definition in place of DAJ's slightly more general one (as described
in their reply to Physical Review Letters (PRL) but not clearly given in DAJ) 
because it is simpler and a counterexample to (7) using Wiseman/Milburn's
more restrictive definition is also a counterexample to (7) under DAJ's 
more inclusive version.

However, because there is no limit involved in Wiseman/Milburn's 
definition (nor in the only description of DAJ's definition given in PRL),
this still leaves DAJ's ``minimal disturbance limit'' partially undefined.
I shall complete the definition by assuming that the ``limit'' refers 
to my original guess \re{eq115}.  In their reply to PRL, the authors
of DAJ refer to 
\re{eq115} as defining ``ideally weak measurement''.  Since I do not know
if they make any distinction between ``ideally weak measurement'' and 
mere ``weak measurement''% 
\footnote{The authors of DAJ have ignored a direct 
question about this, as they have ignored all my recent correspondence.}
\footnote{Our counterexample which assumes what DAJ call 
``ideally weak measurement'' \re{eq115} 
will also be a counterexample under any weaker assumption, which 
presumably includes the mere weak measurement or ``weak limit'' 
assumption of DAJ.  It is annoying that one has to guess at the meaning
of such terms. 
}
I shall use the simpler term ``weak measurement'' to refer to a measurement
satisfying \re{eq115}.%
Our goal is to find a set $\{M_j (g) \}$ of positive measurement operators,
contextual values $\alpha_j (g)$,  
an initial state $\rho$, and a final pure state $f$, 
such that (7) does not hold, assuming the ``weak measurement''
condition \re{eq115}.

We start with equation \re{eq135} for positive measurement
operators $M_j$, but for simplicity we abandon 
DAJ's baroque notation in which \re{eq115} is written 
(e.g., we write $M_j$ in place of $\dajMj$).   
Since a positive operator on a complex Hilbert
space is automatically Hermitian, we replace
all $M^\dag_j$ in \re{eq135} by $M_j$ and explicitly introduce the
weak measurement parameter $g$, obtaining: 
\begin{eqnarray}
\lbl{eq400}
\lefteqn{
\mbox{numerator of (6)} =
}
&& \nonumber \\
&&\tr [ P_f \sum_j \alpha_j 
(1/2)\{M_j(g)  M_j (g) \rho + \rho M_j(g)  M_j(g) \} 
]
\nonumber \\
&&
+  
\tr [P_f \sum_j \alpha_j 
(\  [M_j(g), \rho]\,  M_j(g) 
\  + \  M_j(g)\, [\rho,  M_j(g)]\ ) \, ].
\end{eqnarray}
As noted after \re{eq115}, the first term of \re{eq400} yields (after 
division by the denominator of (6), \re{eq122}) 
the weak value
(7), so a counterexample will result if we can find positive $M_j (g)$ 
satisfying \re{eq115} and  states $\rho, f$ with $\tr [ P_f \rho ] \neq 0$
for which the limit as $g \goesto 0$ of the second term of \re{eq400}  
is nonzero.  This second term can  be rewritten with a double commutator:%
\footnote{I am indebted to a private communication from the authors of 
DAJ for this observation, 
which was part of an attempted proof that (6) implies (7)
in the ``minimal disturbance limit''.  Though not strictly necessary for 
the present counterexample, it provides helpful motivation and simplifies
the calculations.} 
\begin{eqnarray}
\lbl{eq410}
\lefteqn{ 
\tr [P_f \sum_j \alpha_j 
(\  [M_j(g), \rho]\,  M_j(g) 
\  + \  M_j(g)\, [\rho,  M_j(g)]\ ) \, ]   
}
&&
\nonumber
\\ 
&=&\tr [P_f \sum_j - \alpha_j 
 [M_j(g),\, [M_j(g), \rho]\ ]\ ] 
\q.
\end{eqnarray}
If we expand $M_j (g)$ in a power series $ M_j (g) = M^{(0)}_j + 
g M^{(1)}_j + g^2 M^{(2)}_j \ldots $, intuitively 
we would expect the constant term $M^{0}_j$ to be 
a multiple of the identity in order to satisfy the weak measurement condition
\re{eq115}.  Therefore, if $M^{(1)}_j \neq 0$, we would expect 
the double commutator to be of order $g^2$, at least for some states $\rho$. 
Hence to make \re{eq410} nonzero, it would probably be necessary 
to arrange that $\alpha_j (g)$ go to infinity as $1/g^2$ as $g \goesto 0$.   
So, this will be our initial goal, and after achieving it, we shall return to 
analyze \re{eq410} more closely.

Besides assuring that \re{eq410} be nonzero in the limit $g \goesto 0$,
we must define the $M_j (g)$ as positive  operators
such that for some $\{ \alpha_j \}$, the following two equations hold:
\beq
\lbl{eq420}
\sum_j M_j^\dag M_j = \sum_j M^2_j = I
, \mbox{and}
\eeq
\beq 
\lbl{eq430}
\sum_j \alpha_j M^2_j = A 
\q,
\eeq
where $A$ is the operator to be ``weakly measured''.  We shall take 
$A$ to be a one-dimensional projector on a two-dimensional Hilbert space,
represented by a matrix
\beq
\lbl{eq440}
A := 
\left[
\begin{array}{rr}
1 & 0 \\
0 & 0
\end{array}
\right]
\q.
\eeq 
We shall use three measurement operators:
\begin{eqnarray}
\lbl{eq450}
&& M_1 (g) := 
\left[
\begin{array}{cc}
1/2 + g & 0 \\
0 & 1/2 - g
\end{array}
\right]
,\q
M_2 (g) := 
\left[
\begin{array}{cc}
1/2 - g  & 0 \\
0 & 1/2 + g 
\end{array}
\right] ,
\\
&& M_3 (g) := [ I - M^2_1(g) - M^2_2 (g)]^{1/2} =  
\left[
\begin{array}{cc}
\sqrt{1/2  - 2 g^2} & 0 \\
0 & \sqrt{1/2 - 2 g^2} 
\end{array}
\right]
.
\nonumber
\end{eqnarray}
The thing to notice 
is that we are choosing
$M_1$ and $M_2$ small enough that $M_3$ is uniquely determined by \re{eq420}
and in such a way that some of the contextual values $\alpha_j (g)$ 
can be chosen to be of order $1/g^2$. 

Writing out \re{eq430} in components gives two scalar equations in 
three unknowns:
\begin{eqnarray}
\lbl{eq460}
(1/2 + g)^2 \alpha_1 (g) + (1/2 - g)^2 \alpha_2 (g) + 
(1/2 - 2g^2) \alpha_3 (g) &=& 1 \\
(1/2 - g)^2 \alpha_1 (g) + (1/2 + g)^2 \alpha_2 (g)  
+ (1/2 - 2g^2) \alpha_3 (g) &=& 0 \nonumber \q.
\end{eqnarray}
Since we want to make at least one of the $\alpha_j$ of order  
$1/g^2$, let us attempt to define $\alpha_1 (g) := 1/g^2$ and solve
the remaining system for $\alpha_2$ and $\alpha_3$.  The remaining system
is consistent, and the full solution is:
\beq
\lbl{eq470}
\alpha_1 (g) := \frac{1}{g^2}, \q 
\alpha_2 (g) = \frac{1}{g^2} - \frac{1}{2g}, \q
\alpha_3 (g) = - \frac{4g^3 - 12g^2 + g - 4}{4g^2(4g^2-1)}
\q.
\eeq
(This has been checked both by hand and by a computer algebra program.)

To see with minimal  calculation that this will produce a counterexample,
note that for  
\beq
\lbl{eq480}
\rho = 
\left[
\begin{array}{cc}
\rho_{11} & \rho_{12}\\
\rho_{21}   & \rho_{22}
\end{array}
\right]
\eeq 
and for any diagonal matrix 
$$
D = 
\left[
\begin{array}{cc}
d_{1} & 0\\
0   & d_2
\end{array}
\right]
\q,
$$
$$
[D, \rho] = 
\left[
\begin{array}{cc}
0  & (d_1 - d_2) \rho_{12}\\
(d_2 - d_1)\rho_{21}   & 0 
\end{array}
\right], 
\  \mbox{and} \  
$$
$$
[D, [D, \rho]] = 
\left[
\begin{array}{cc}
0  & (d_1 - d_2)^2\rho_{12}\\
(d_2 - d_1)^2 \rho_{21}   & 0
\end{array}
\right]
.
$$
In particular for $j = 1,2$,
$$
[\ M_j(g),\, [M_j(g), \rho]\,] =  
\left[
\begin{array}{cc} 
0 & 4g^2 \rho_{12} \\
4g^2 \rho_{21} &  0
\end{array}
\right] \q,
$$
and since $M_3 (g)$ is a multiple of the identity, $[M_3(g), \rho] = 0.$
Hence \re{eq410} becomes:
\begin{eqnarray}
\lbl{eq490}
\lefteqn{(1/2)\tr [P_f \sum_j - \alpha_j 
 [M_j(g),\, [M_j(g), \rho]\ ]\ ] =}\\ 
&& - \tr [P_f 
\left[
\begin{array}{cc}
0 & 4\rho_{12}\\
4\rho_{21}   & 0\\ 
\end{array}
\right] ] + O(g) 
.
\end{eqnarray}
The trace is easily seen to be nonzero for $\rho_{12} \neq 0$ and appropriate
$P_f$ (see below).

Combining \re{eq120}, \re{eq122}, and  
\re{eq410}, gives the weak limit 
of (6) for this example as expression (7) 
plus a term which does not always vanish.
For example, for a norm 1 vector $f := 
(f_1, f_2) $  
\beq
\lbl{eq495}
\mbox{weak limit of (6)} = 
\frac{\tr [ P_f \{A, \rho \} ]}
{2 \tr [P_f \rho ]} 
+ \frac{- 8 \Re ({f}^*_2 f_1 \rho_{21}) }
{|f_1|^2 \rho_{11}  + 2 \Re ({f}^*_2 f_1 \rho_{21}) + |f_2|^2 \rho_{22} }.  
\eeq
The first term is DAJ's expression (7) written in our notation. 
\begin{eqnarray}
\lbl{eq500} 
\lefteqn{\mbox{difference between weak limit of (6) and expression (7)} =}
&&
\nonumber\\ 
&&
\frac{- 8 \Re ({f}^*_2 f_1 \rho_{21}) }
{|f_1|^2 \rho_{11}  + 2 \Re (f^*_2 f_1 \rho_{21}) + |f_2|^2 \rho_{22} }.  
\end{eqnarray}

Some may feel uneasy about this example because $M_3 (g)$ is a multiple
of the identity, which may seem suspiciously trivial.%
\footnote{This example may be somewhat artificial, 
but it is not physically unrealistic.
One could realize this situation by first performing a classical experiment
with two outcomes called $A$ and $B$, each occuring with a definite 
probability.  If $A$ occurs, perform a quantum experiment with outcomes 
1 and 2; if $B$ occurs, do nothing further.  Outcome $B$ would correspond
to outcome 3 in a quantum experiment with measurement operators $M_1,
M_2, M_3$.
} 
The above measurement operators were chosen to give a simple example
with minimal calculation. 
Less trivial examples should be obtainable by appropriately perturbing the
above measurement operators, e.g., 
adding to $M_1$ or $M_2$ terms 
which are of order $g^2$ or greater, so that in the limit $g \goesto 0$
they will not affect the anomalous term in \re{eq400}.  
I have not pursued this because the calculations rapidly get messy, 
and it is not clear what would be learned from them. 
\section{Appendix: 
A mathematician's view of the Moore-Penrose pseudoinverse }
It took me perhaps half an hour to translate DAJ's complicated description 
of the Moore-Penrose pseudoinverse into something which I could 
easily visualize.  Such descriptions are not uncommon in the
physics literature.  This appendix is written as a service to those 
who might be interested in how mathematicians think about such things.

Let $A$ be an operator from a Hilbert space $S$ to
a possibly different Hilbert space $K$.  For simplicity, I will assume
below that all Hilbert spaces mentioned are finite dimensional, but
nearly all of what will be said also works for infinite dimensional spaces,  
with appropriate changes of language and qualifications, 
Many people feel more comfortable
thinking of $A$ as a matrix, but mathematicians have learned that unless
the problem at hand specifically involves matrices, it is usually easier
and more insightful not to.

There are three important subspaces of associated with a given
operator $A$.  Its {\em nullspace}, denoted $\nullsp (A)$ is defined
as the set of all vectors $n \in S$ such that $An = 0$.  Its {\em range},
denoted $\range (A)$ is defined as the set of all vectors $r$ in $K$
which are of the form $r =As$ for some vector $s \in S$.  Perhaps less
well known is the {\em initial space} of $A$, denoted $\init (A)$, which
is defined as the orthogonal complement of the nullspace:
$$
\init(A) := \nullsp (A)^\perp := \{s \in S |\ \lb s, n \rb = 0 
\q \mbox{for all  $n \in \nullsp (A)$} \}.  
$$

The restriction of $A$ to its initial space, denoted $A | \init(A)$,
maps $\init(A)$ onto $\range(A)$, and {\em this restriction is always
1:1}.  This is because if $A|\init(A): \init(A) \rightarrow K$ 
were not 1:1,
it would have a nontrivial nullspace, but $\init(A)$ is, by definition,
orthogonal to $\nullsp(A)$.  
%(One also has to check that $\init(A)$ maps 
%{\em onto} $\range(A)$, which follows immediately from the fact that 
%$S is a direct sum of $\nullsp(A)$ and $\nullsp(A)^\perp =: \init(A)$). 

This suggests that it might be profitable to temporarily change our 
object of study from the original operator
$A: S \rightarrow K$ to a new operator $A| \init (A) :  
\init(A)\rightarrow K$.  
In accordance with this broader view, 
consider an operator 
$C: H \rightarrow K$ which maps a Hilbert space $H$ into a Hilbert space 
$K$.  (We do not assume that $C$ maps $H$ {\em onto} $K$.)  

An operator $B: K \rightarrow H$ is called a {\em left inverse } of 
$C: H \rightarrow K$ if $BC = I_H$, where $I_H$ denotes the identity operator
on $H$.  It is easy to describe how to construct all left inverses
$B$ for $C$.  Any left inverse is uniquely determined on the 
range of $C$, i.e., for 
$k = Ch \mbox{ with } h \in H$,
by the equation  
$$
Bk = BCh = I_H h = h \q.
$$ 
If given $k \in K$ there is more than one $h \in H$ with $Ch = k$
(i.e., if $C$ is not 1:1), 
then $C$ has no left inverse.   

That uniquely defines the left inverse $B$ on $\range(C)$
when $C$ is 1:1.
$B$ can be defined any way we choose (subject to linearity) 
on $\range(C)^\perp$.
Since $K$ is the direct sum of $\range(C) $ and $\range(C)^\perp$,
$B$ is uniquely defined on all of $K$ once it is defined on $\range(C)$ and 
$\range(C)^\perp$.

The simplest way to define $B$ on $\range(C)^\perp$ is to make it identically
zero there, and this leads us to the definition of the 
Moore-Penrose pseudoinverse, denoted $A^+$,
for an operator
$A: S \rightarrow K$.
We just take  $H := \init (A)$ and apply the preceding discussion.
Define $A^+ : K \rightarrow \init(A) \subset S$ to be the unique left inverse 
for $A|\init(A)$
which is zero on $\range(A)^\perp$.
If one works through the complicated definition of the 
Moore-Penrose pseudoinverse given in DAJ, 
 one sees that this is how it is defined, at least for the case in 
which the eigenspaces of $A A^\dag $ corresponding to nonzero eigenvalues
are one-dimensional.%
\footnote{For the general case, the prescription of DAJ is either 
incomplete or incorrect, depending on the interpretation.  The question is
the meaning of their definition of $\dajU$ as the orthogonal matrix 
``composed of the eigenvectors of $\dajF \dajF^\dag$''.  I would guess
that this would mean that the columns of $\dajU$ are eigenvectors of 
$\dajF  \dajF^\dag$ (which still only defines $\dajU$ 
up to multiplying its columns by scalars when the eigenvalues of 
$\dajF \dajF^\dag$ all have multiplicity 1, with further ambiguity 
possible in more general cases),
but if that is what they meant, then their construction
is incorrect.  A simple counterexample is a unitary matrix $\dajF$ which
is not diagonal, in which case $\dajU, \dajV,$ and ${\bf \Sigma}$ 
could all be diagonal under the above interpretation.
}%%%

Most mathematicians probably would define
the Moore-Penrose pseudoinverse $A^+$ of $A$ as the unique left inverse  
for $A|\init(A): \init(A) \rightarrow K$ 
which annihilates the orthogonal complement of $\range(A)$.
It is a genuine left inverse for $A | \init(A) : \init(A) \rightarrow K$,
but not for $A: S \rightarrow K$;  the latter has no left inverse
unless $\init(A) = S$ because otherwise $A$ has a nontrivial nullspace 
and so is not 1:1.  

That is concise and easy to visualize.  In addition, it generalizes
immediately to operators $A$ on an infinite-dimensional Hilbert space 
space  (without the compactness hypothesis for $A$ 
mentioned in footnote 12 of DAJ), in which case the
pseudoinverse will be a so-called ``closed'', possibly unbounded, operator.
This pseudoinverse will be bounded if and only if
0 is an isolated point of the spectrum of $A$. 
(Even if $A$ is compact, this pseudoinverse may be unbounded.)

If one is given $A$ as a matrix, the question of how to write down 
the Moore-Penrose pseudoinverse as a matrix naturally arises.  This is 
a computational, rather than conceptual, question which is a bit messier, 
and that is where the singular value decomposition
may come 
in (though one can avoid it by using the simpler polar decomposition).   
If one is dealing with matrices, one may be forced into messy matrix 
descriptions, but otherwise it is usually profitable to avoid them.

\end{document}